# Oxygen vacancy formation in ZnSeTe blue quantum dot light-emitting diodes

*Shaun Tan,[1,†] Sujin Park,[2,3,†] Seung-Gu Choi,[4,†] Oliver J. Tye,[1] Ruiqi Zhang,[6] Jonah R. Horowitz,[1] Heejae Chung,[2] Vladimir Bulović,[6] Jeonghun Kwak,[3] Jin-Wook Lee,[4,5,*] Taehyung Kim,[2,*] Moungi G. Bawendi[1,*]*

[1]Department of Chemistry, Massachusetts Institute of Technology, Cambridge, MA, USA

[2]Samsung Advanced Institute of Technology, Samsung Electronics, Suwon-si, Republic of Korea

[3]Department of Electrical and Computer Engineering, Inter-University Semiconductor Research Center and Soft Foundry Institute, Seoul National University, Seoul, Republic of Korea

[4]Department of Nano Engineering and Department of Nano Science and Technology, SKKU Advanced Institute of Nanotechnology (SAINT), Sungkyunkwan University, Suwon 16419, Republic of Korea.

[5]SKKU Institute of Energy Science & Technology (SIEST), Sungkyunkwan University, Suwon, 16419, Republic of Korea

[6]Department of Electrical Engineering and Computer Science, Massachusetts Institute of Technology, Cambridge, MA, USA.

†These authors contributed equally to this work.

*Email: jw.lee@skku.edu, thyun9.kim@samsung.com, mgb@mit.edu

**ABSTRACT**

Recent advancements have led to the development of bright and heavy metal-free blue-emitting quantum dot light-emitting diodes (QLEDs). However, consensus understanding of their distinct photophysical and electroluminescent dynamics remains elusive. This work correlates the chemical and electronic changes occurring in a QLED during operation using depth-resolved and operando techniques. The results indicate that oxygen vacancy forms in the ZnMgO layer during operation, with important implications on the charge injection and electrochemical dynamics. Taken together, the results suggest a causal relationship between oxygen vacancy formation and operational degradation of the blue-emitting ZnSeTe-based QLEDs.

**INTRODUCTION**

Electroluminescent quantum dot light-emitting diodes (QLEDs) are strong contenders for future advanced display applications, offering unique advantages such as a wide color gamut, high spectral purity, and low-cost solution processability.[1–3] However, the subpar device characteristics of blue-emitting QLEDs continue to hinder their practical adoption. Compared to red- and green-emitting devices, blue QLEDs typically exhibit lower external quantum efficiency, shorter operational lifetimes, reduced spectral purity, and more pronounced positive aging behavior.[4–7] These challenges are often more severe for the heavy-metal free blue emitters based on ZnSeTe quantum dots.[4,8,9] The wide bandgap of blue quantum dots creates distinct interfacial energy alignments with adjacent functional layers in a QLED, which is expected to give rise to unique charge injection dynamics that are not experienced by red or green QLEDs. However, the fundamental mechanisms by which the large bandgap of blue emitters influences charge balance, exciton recombination, and device degradation remain insufficiently understood, slowing progress and development in the research field.

This work investigates the chemical and electronic changes in a blue ZnSeTe QLED. We combine spatially- and depth-resolved characterization techniques to study the various functional layers and interfaces of a complete QLED, and observed a decrease in the intrinsic work function of certain functional layers that occurred in a sequential manner. We show that the changes are associated with oxygen vacancy defects forming during operation, with important consequences on the charge injection, electrochemical dynamics, and operational degradation of the QLEDs.

**RESULTS AND DISCUSSION**

A representative high-resolution transmission electron microscopy (TEM) image (**Figure 1**a) shows the QLED device structure. The QD emissive layer is composed of blue-emitting ZnSeTe/ZnSe/ZnS core/shell/shell QDs, which will henceforth be simply referred to as ZnSeTe. This work is not related to device optimization so the performance parameters (Figure S1) are briefly discussed here. The device exhibits pure blue electroluminescence with a 465 nm peak and 36 nm full width half maximum. The external quantum efficiency and brightness were up to 16.2% and 61,000 cd m$^{-2}$, respectively, and the $T_{50}$ lifetime was 16,683 hours at 100 cd m$^{-2}$ based on an empirical acceleration factor of 1.8.

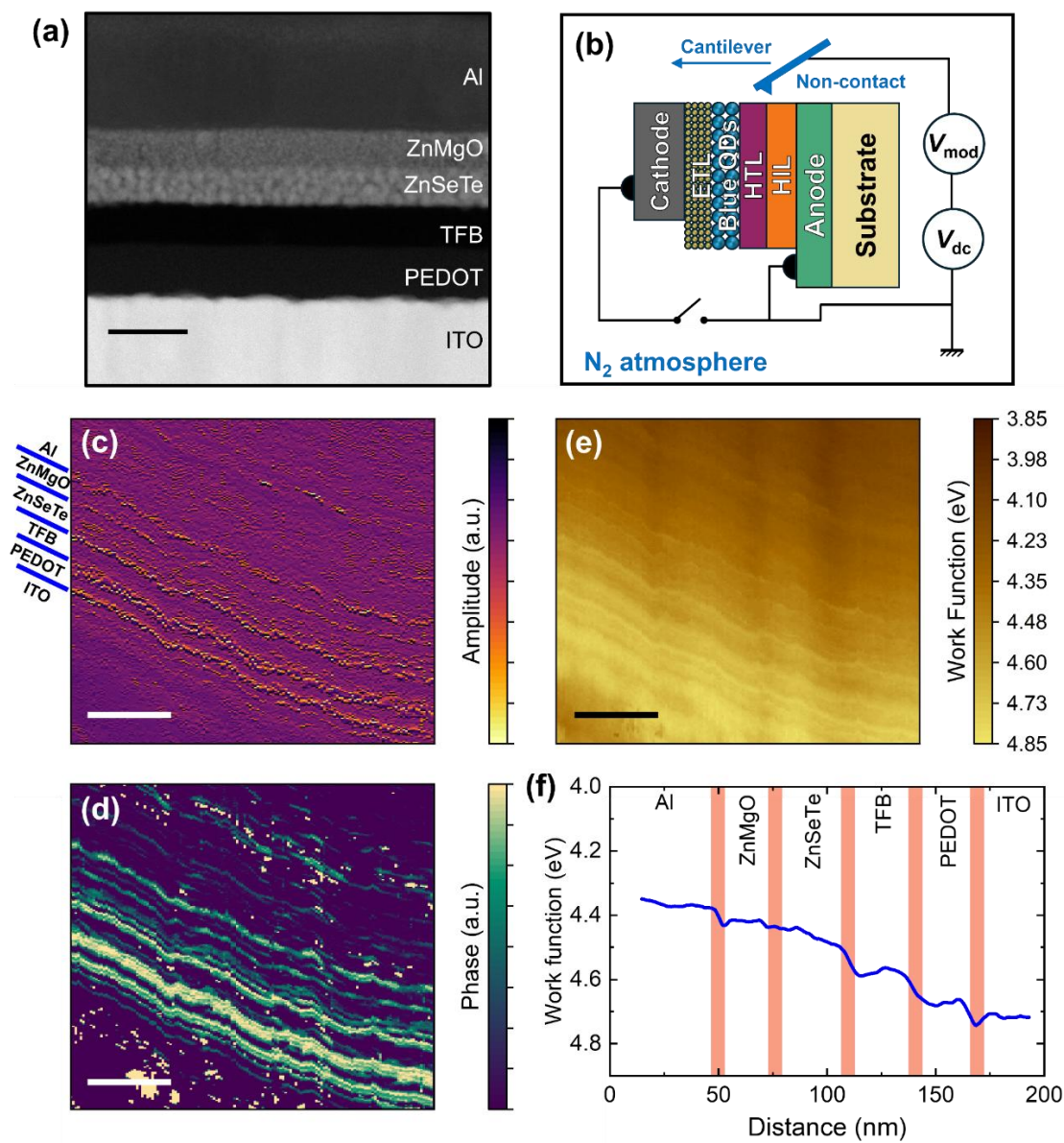

**Figure 1.** (a) Cross-sectional TEM high-angle annular dark-field (HAADF) image of the QLED structure. Scale bar is 50 nm. (b) Schematic of the non-contact mode KPFM measurement setup performed on a complete QLED device in an inert nitrogen atmosphere. Oscillation (c) amplitude and (d) phase shift mapping of the QLED device cross-section. (e) Work function map and (f) average work function distribution of the QLED. Scale bars in (c)-(e) are 100 nm.

Single-pass non-contact mode Kelvin Probe Force Microscopy (KPFM) (Figure 1b) was used to profile the electronic properties of a QLED cross-section. For our experiments, we note that focused ion beam (FIB) milling was *not* used to polish the QLED cross-section for the KPFM measurements, because FIB is well-known to damage and unintentionally interfere with the electronic properties of a sample surface.[13–15] The probe oscillation amplitude (Figure 1c) and phase shift (Figure 1d) are useful to distinguish the different device layers. As the cantilever rasters across the different functional layers, the oscillation frequency and amplitude are changed by the different surface physical and mechanical interactions within the non-contact regime while the feedback loop readjusts the modulating voltage on the cantilever to maintain a constant oscillation frequency and amplitude. In practice this means that the amplitude and phase contrast can be used to identify the layer interfaces. Subsequently, the greatest strength of non-contact mode KPFM is its ability to non-destructively map the work function spatial distribution across the entire device stack of a complete QLED (Figure 1e, 1f, and Figure S2). The mapping results show that the work function becomes progressively larger (i.e. more p-type) going from the Al cathode to the ITO anode, consistent with the electron and hole injection directions. Band bending occurs at the interfaces between adjacent functional layers due to the electric field created by a depletion region. All interfaces generally show obvious band bending, except the minor offset at the ZnMgO/QD interface, suggesting that the work function of the ZnMgO ETL and ZnSeTe QDs are relatively comparable. More broadly, while ultraviolet photoelectron spectroscopy (UPS) can measure the work function of individual materials, it lacks spatial and depth resolution and is unable to characterize the full work function distribution across the complete QLED device. In particular, UPS measurements cannot probe interface effects. To the best of our knowledge, we are reporting the first measurement of the work function spatial distribution in a complete QLED.

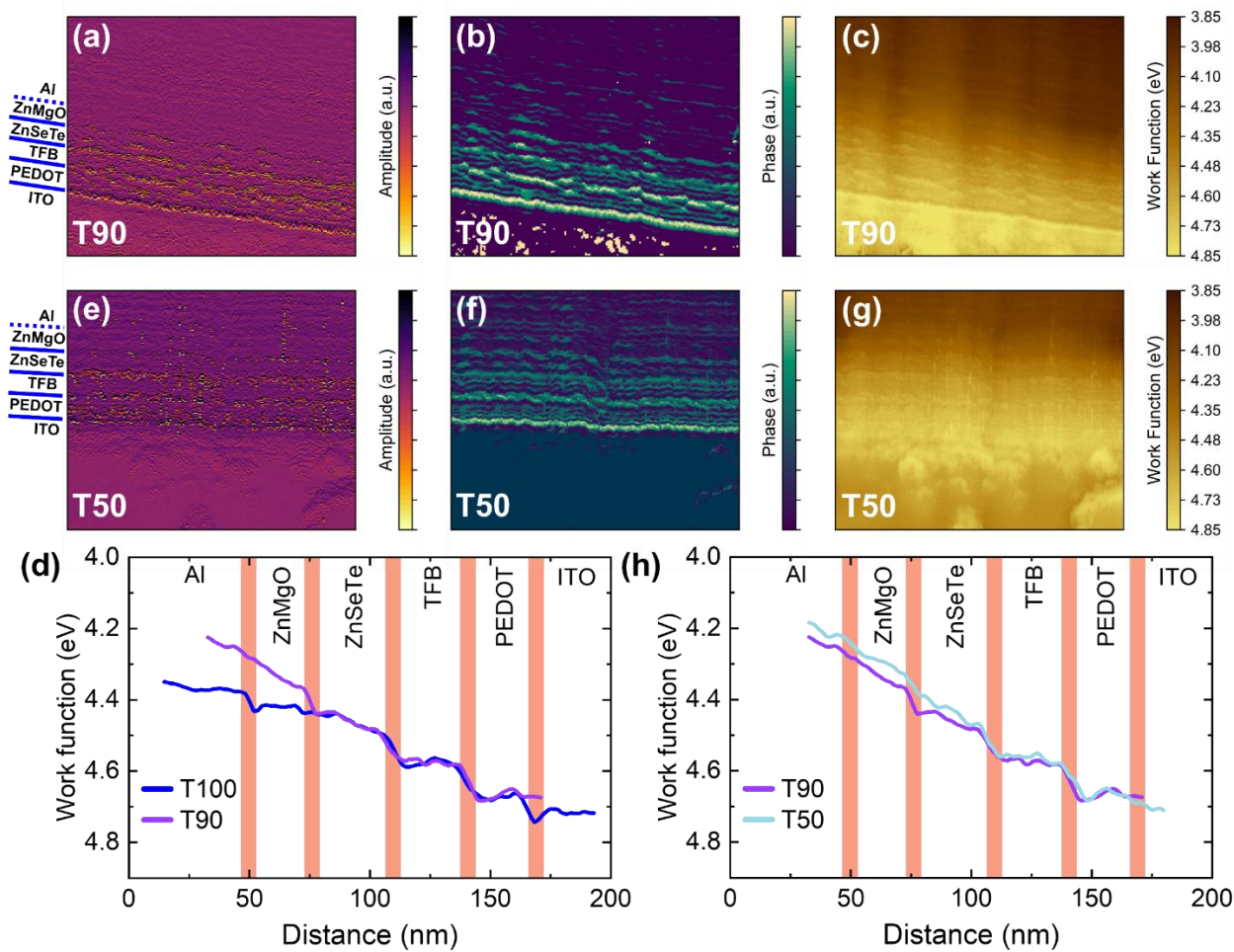

**Figure 2.** Oscillation (a) amplitude, (b) phase shift, and (c) work function mapping of the QLED at its T90 lifetime. Oscillation (d) amplitude, (e) phase shift, and (f) work function mapping of the QLED at its T50 lifetime. Comparison of the average work function distribution of the QLED (g) between T100 (pristine) and T90, and (h) between T90 and T50.

We then proceeded to study how the QLED evolves during degradation. Operando KPFM was performed on the same cross-sectional area (Figure S3) to directly compare how the various functional layers and interfaces change before and after QLED operation. We note that the operando KPFM was performed entirely in an inert nitrogen glovebox to eliminate the influence of ambient water and oxygen. Henceforth the results previously discussed in Figure 1 for the pristine device before aging are referred to as T100.

After constant current operation to T90 (QLED after brightness dropped to 90% of its initial value), the work function of the ZnMgO ETL was upshifted (more n-type behavior) relative to the pristine

T100 case (**Figure 2**a-2d). In contrast, the ZnMgO/QD interface remained unchanged, and likewise the work function distributions across the ZnSeTe QDs, TFB, and PEDOT:PSS were mostly the same at T90. The Al/ZnMgO interface boundary, which at T100 was clearly distinguishable, had become less defined in the oscillation amplitude and phase shift maps at T90 (dashed blue line in Figure 2a). This suggests degradation of the Al/ZnMgO interface possibly related to species diffusion and layer intermixing which will be discussed later.

After subsequent aging to T50 (QLED after brightness dropped to 50% of its initial value), the ZnMgO ETL further became more n-type (Figure 2e-2h), but the corresponding decrease in work function from T90 to T50 was comparatively modest relative to the initial upshift from T100 to T90. Focusing on the ZnSeTe emissive layer and ZnMgO/QD interface, between T100 and T90 there was no significant difference, but between T90 and T50 the region closer towards the ZnMgO/QD interface (Figure 2h) showed a decreased work function, whereas the region adjacent to the TFB HTL was still comparatively similar. This trend suggests that the change is propagating inwards from the ZnMgO/QD interface. The sequential order of events between T100, T90, and T50 in addition to the propagating trend imply that the ZnMgO was the cause that initiated the effects observed between T90 and T50 for the ZnSeTe QD layer. Notably the work function profiles of the TFB HTL and PEDOT:PSS HIL were relatively unchanged going from T100, T90, to T50.

There is active debate in the research community regarding which specific functional layer (or combination of multiple layers) and failure mechanisms are responsible for the short operational lifetimes of blue QLEDs. A popular notion from recent blue QLED literature[5,16–18] is that changes associated with the organic HTL and HIL (e.g. electron leakage, structural deformation, defect formation) are to be blamed. Organic polyfluorenes such as TFB are understood to be vulnerable to electrochemical reactions induced by leakage electrons to form trap states and deteriorate its hole conductivity.[19–21] However, our results indicate that the TFB HTL likely did not *initiate* the QLED degradation. Meanwhile, any type of electrochemical redox reaction happening to TFB would have changed its charge state and hole density and led to a measurable change in work function. We do not suspect that this is the case since the work function of TFB was comparable between T100 and T50. Separate literature suggested ligand detachment from the QD surface[16,22,23] as the culprit responsible for the instability of blue QLEDs. Loss of capping ligands leaves behind unpassivated surface traps that serve as non-radiative recombination centers that reduce the quantum yield of the emissive layer. Perhaps this is related to the changes in the ZnSeTe QDs we observed between T90 and T50, but it does not explain the prior lifetime drop from T100 and T90. Additionally, ligand detachment would still be an effect, and the ZnMgO ETL remains the initiating cause.

Suspecting that the observed n-doping of ZnMgO is electrochemical in nature, we proceeded to study its chemical environment using X-ray photoelectron spectroscopy (XPS). We continued to prioritize our goal of studying the complete functional QLED rather than isolated thin films of individual layers, so we had to etch away the overlying Al cathode of the QLED device using an argon beam to reach the buried ZnMgO ETL (**Figure 3**a, 3b). We first examined the chemical states of Al in the bulk (~70 nm deep). No significant difference is observed for the Al 2s (~118.7 eV) and O 1s (~531.4 eV) peak positions between the pristine (T100) and aged (T50) QLEDs (Figure 3c). However, at T100 the O:Al ratio was 0.42, which increased to 0.54 for the QLED at T50, indicating a higher oxygen content in the Al cathode. The oxygen may have originated from chemical diffusion and intermixing of oxygen from the ETL. We can exclude batch-to-batch variations from causing the difference in the oxygen content because both the T100 and T50 device pixels were adjacent to one another on the same substrate (Figure S4). This suggests that oxygen diffusion may have occurred in the aged QLED.

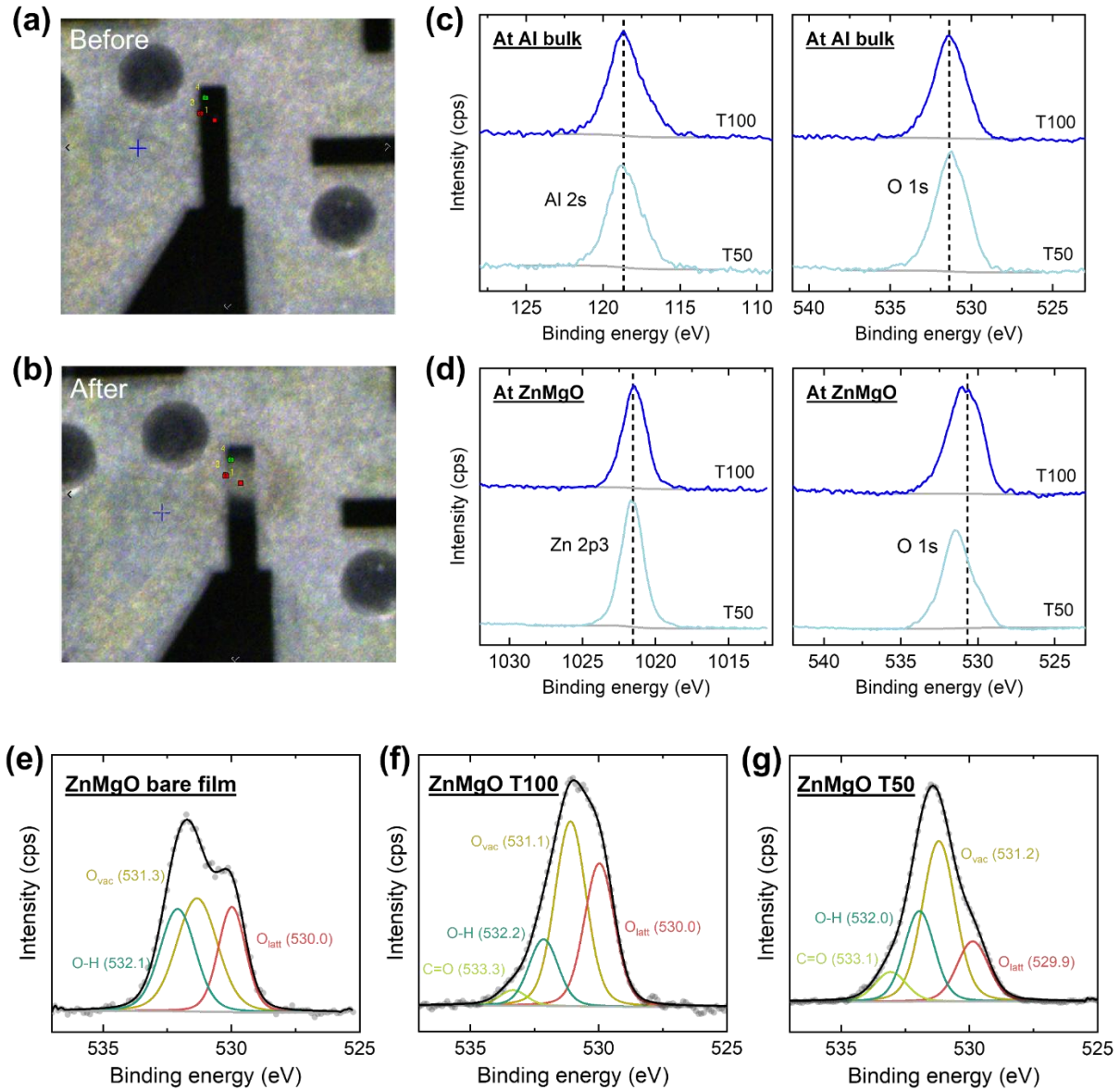

**Figure 3.** Optical microscopy images of the QLED device (a) before and (b) after argon etching to reach the buried ZnMgO ETL. (c) High-resolution XPS spectra of the Al 2s (left) and O 1s peaks (right) of the Al cathode bulk. (d) High-resolution XPS spectra of the Zn 2p3 (left) and O 1s peaks (right) of the ZnMgO ETL. Deconvolution and fitting of the O 1s peak for the (e) bare film of ZnMgO, (f) pristine (T100) ZnMgO in a QLED, and (g) degraded (T50) ZnMgO in a QLED. The fit lines are given by the black solid lines.

For direct evidence, we further etched the Al cathode to reach the ZnMgO ETL. The scotch tape peel-off method on the cathode was avoided to maintain thickness uniformity and to prevent damage to the thin layers and interfaces. Inspecting the high-resolution core level spectra of ZnMgO, the Zn 2p3 peak position at ~1021.5 eV (Figure 3d) was mostly similar between the pristine and degraded QLEDs. In contrast, the O 1s spectra showed a different peak shape profile at T50 with an asymmetric skew towards higher binding energy. Both O 1s spectra had broadened peak shapes with noticeable asymmetric tails, indicating the presence of multiple chemically distinct oxygen states. We also characterized the bare ZnMgO film (Figure S5) directly spincoated on a glass substrate for comparison. Deconvolution of the O 1s spectra (Figure 3e-3g) allows us to distinguish the constituent peaks of the lattice oxygen at ~530 eV ($O_{latt}$) and the oxygen associated with vacancies at ~531 eV ($O_{vac}$).[24–27] For the latter, we note that an oxygen vacancy defect (i.e. the absence of an oxygen atom) does not directly emit photoelectrons, but its presence is inferred from the higher binding energy at approximately 531 eV[24–27] of remaining oxygen atoms neighboring the vacancy due to the altered chemical bonding environment, which is what we are referring to as $O_{vac}$. One report assigned the peak at ~531 eV to chemisorbed $H_2O$ based only on theoretical calculations.[28] However, experimental evidence is lacking, and that interpretation is not generally accepted in the community.[24–27] We thus correlate the peak at ~531 eV to oxygen vacancies in this study. The ZnMgO of the pristine QLED had a $O_{vac}$:$O_{latt}$ ratio of 1.34 (**Table 1**), slightly smaller than the $O_{vac}$:$O_{latt}$ ratio of 1.59 for the ZnMgO bare film. In contrast, the degraded QLED at T50 had a noticeably larger $O_{vac}$:$O_{latt}$ ratio of 2.79, providing clear evidence of lattice oxygen loss and the formation of additional oxygen vacancy defects. This links back to the diffusion and increased concentration of oxygen in the Al cathode. This is also consistent with the less defined ETL/QD interface observed during the operando KPFM mapping as previously discussed. Put simply, oxygen vacancies are forming in the ZnMgO ETL during QLED degradation.

**Table 1.** Deconvolution of the O 1s spectra for ZnMgO in the pristine and degraded QLEDs

|  | ZnMgO bare film | | Pristine T100 | | Degraded T50 | |
| --- | --- | --- | --- | --- | --- | --- |
|  | Binding energy (eV) | Atomic percent (%) | Binding energy (eV) | Atomic percent (%) | Binding energy (eV) | Atomic percent (%) |
| $O_{latt}$ | 530.0 | 25.9 | 530.0 | 34.7 | 529.9 | 17.5 |
| $O_{vac}$ | 531.3 | 41.5 | 531.1 | 46.3 | 531.2 | 48.8 |
| O-H | 532.1 | 32.6 | 532.2 | 15.6 | 532.0 | 25.1 |
| C=O | - | | 533.3 | 3.4 | 533.1 | 8.6 |
| $O_{vac}$:$O_{latt}$ | 1.59 | | 1.34 | | 2.79 | |

Tying together all preceding discussions, we propose that the formation of oxygen vacancies is linked to the instability of blue QLEDs. In typical metal oxide semiconductors, oxygen vacancies are n-type donor defects because the unbonded electrons delocalize into the metal oxide conduction band and the free electron density is resultingly increased.[29,30] The same applies to ZnO or ZnMgO,[31–33] where a loss of an oxygen atom generates two free electrons according to:

$$O_O^\times \rightarrow V_O^{\cdot\cdot} + 1/2 O_2 + 2e^-$$

where $V_O^{\cdot\cdot}$ refers to an oxygen vacancy defect in Kröger–Vink notation. Oxygen vacancy formation therefore explains the upshifted work function (i.e. more n-type behavior) of the ZnMgO ETL observed during QLED degradation. Band diagrams constructed based on the KPFM results illustrate the reduced electron injection barrier caused by the upshifted work function (Figure S6). The reduced injection barrier is also consistent with the lower turn-on voltage and higher leakage current of the QLED (Figure S7). Overall, oxygen vacancy formation was responsible for the increased free electron density, reduced injection barrier, and higher ZnMgO conductivity, and the consequence is an excess injection of electrons into the ZnSeTe emissive layer.

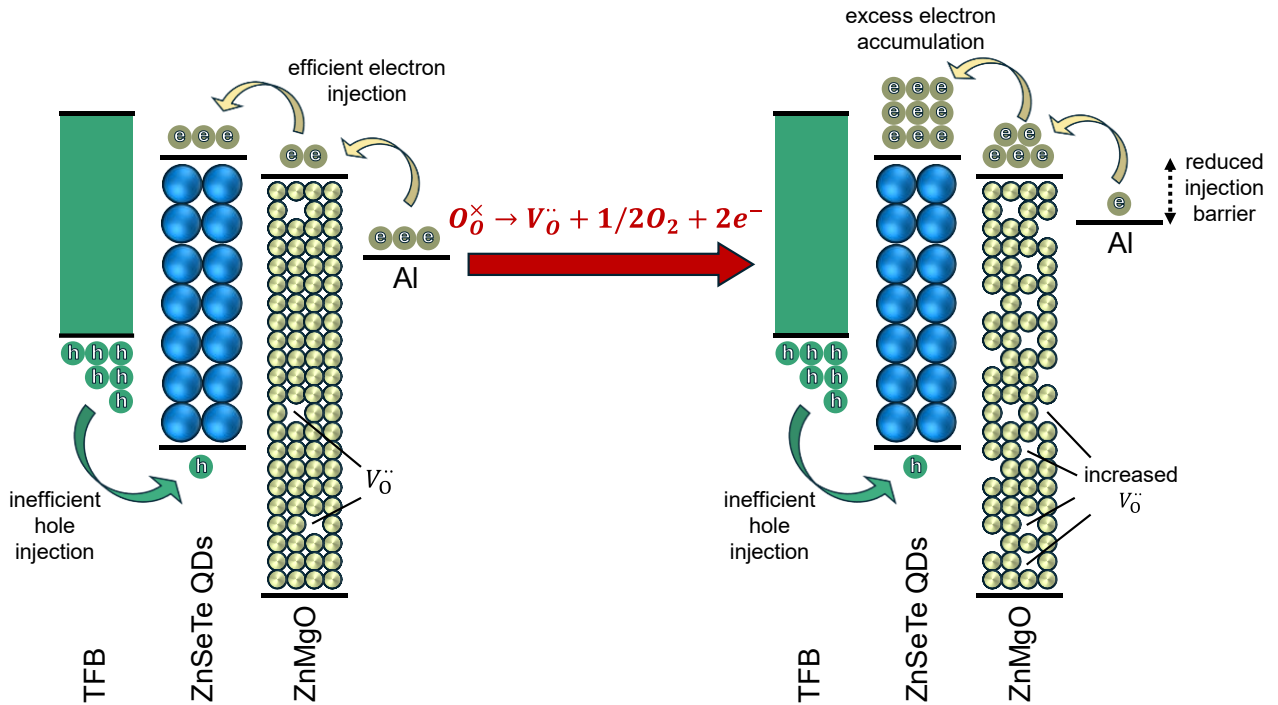

**Figure 4.** Oxygen vacancy formation during operation and its effects on the charge injection dynamics of the QLED.

The wide bandgap and deep valance band level of blue QDs compared to red or green emitters creates a large energy alignment offset at the QD/HTL interface which hinders hole injection into the emissive layer (**Figure 4** and Figure S8).[5,34,35] Additionally, the electron mobility of metal oxide ETLs is typically higher than the hole mobility of organic HTLs.[2,36] Thus, the major challenge with blue QD emitters is that electron injection is already in excess in a pristine QLED. Further increasing electron injection is undesirable, and will worsen the existing charge imbalance and aggravate electron accumulation in the emissive layer. The excess electrons may then trigger or exacerbate a variety of electrochemical processes that cause QLED degradation such as surface ligand desorption from the QD,[22,37] electron leakage into the HTL and HIL and associated redox reactions,[18,20] and non-radiative Auger recombination processes[38] that quench the emissive layer quantum yield. The increased proportion of the O-H (~532.1 eV) and C=O (~533.2 eV) constituent peaks at T50 (Table 1) may have been caused by detached organic ligands (e.g. oleic acid) diffusing to the ZnMgO ETL, which will be the subject of future investigations. Regardless, the formation of oxygen vacancies in ZnMgO was the catalyst that resulted in all such electrochemical degradation processes.

We briefly discuss potential solutions to alleviate the excess electron injection. One possibility is to introduce an electron-blocking layer at the ETL/QD interface to limit electron injection,[39] but this may come at the cost of reduced brightness and current efficiency. Another strategy is to increase hole injection to improve charge balance, either by employing new materials[18] with a smaller energy gap at the QD/HTL interface, or by incorporating interlayers with a stepwise energy alignment[5] to bridge hole transport. More broadly, our work elucidated the link between oxygen vacancy formation and the degradation of ZnSeTe blue QLEDs, which will guide future studies on understanding their electrochemical dynamics and improving their device characteristics.

**SUPPORTING INFORMATION**

The Supporting Information is available free of charge at

- Materials and Methods, QLED performance metrics, KPFM mapping of QLED cross-section, Operando KPFM, Photograph of the XPS sample, XPS of ZnMgO, Band alignments based on the KPFM results, Current-voltage curve change, Blue versus red QDs

**AUTHOR INFORMATION**

**Corresponding Authors**

**Jin-Wook Lee** – *Department of Nano Engineering and Department of Nano Science and Technology, SKKU Advanced Institute of Nanotechnology (SAINT), and SKKU Institute of Energy Science & Technology (SIEST), Sungkyunkwan University, Suwon 16419, Republic of Korea*; Email: jw.lee@skku.edu

**Taehyung Kim** – *Samsung Advanced Institute of Technology, Samsung Electronics, Suwon-si, Republic of Korea*; Email: thyun9.kim@samsung.com

**Moungi G. Bawendi** - *Department of Chemistry, Massachusetts Institute of Technology, Cambridge, MA, USA*; Email: mgb@mit.edu

**Authors**

**Shaun Tan** - *Department of Chemistry, Massachusetts Institute of Technology, Cambridge, MA, USA*

**Sujin Park** - *Samsung Advanced Institute of Technology, Samsung Electronics, Suwon-si, and Department of Electrical and Computer Engineering, Inter-University Semiconductor Research Center and Soft Foundry Institute, Seoul National University, Seoul, Republic of Korea*

**Seung-Gu Choi** - *Department of Nano Engineering and Department of Nano Science and Technology, SKKU Advanced Institute of Nanotechnology (SAINT), Sungkyunkwan University, Suwon 16419, Republic of Korea*

**Oliver J. Tye** - *Department of Chemistry, Massachusetts Institute of Technology, Cambridge, MA, USA.*

**Ruiqi Zhang** - *Department of Electrical Engineering and Computer Science, Massachusetts Institute of Technology, Cambridge, MA, USA*

**Jonah R. Horowitz** - *Department of Chemistry, Massachusetts Institute of Technology, Cambridge, MA, USA.*

**Heejae Chung** - *Samsung Advanced Institute of Technology, Samsung Electronics, Suwon-si, Republic of Korea*

**Vladimir Bulović** - *Department of Electrical Engineering and Computer Science, Massachusetts Institute of Technology, Cambridge, MA, USA*

**Jeonghun Kwak** - *Department of Electrical and Computer Engineering, Inter-University Semiconductor Research Center and Soft Foundry Institute, Seoul National University, Seoul, Republic of Korea*

**Author Contributions**

S.T., S.P., and S.-G.C. contributed equally to this work.

**Notes**

The authors declare no competing financial interest.

**ACKNOWLEDGMENTS**

S.T., O.J.T., J.R.H. and were supported by the Samsung Advanced Institute of Technology (SAIT). This work was supported by SAIT, Samsung Electronics Co., Ltd. This work was performed in part in the MIT.nano Characterization Facilities. S.-G. C. and J.-W. L. acknowledge financial support from the Korea Institute of Energy Technology Evaluation and Planning (KETEP) through a grant funded by the Ministry of Trade, Industry and Energy (MOTIE) of the Republic of Korea (Grant Nos. RS-2023-00266248 and RS-2022-KP002701). The authors would like to thank Prof. Nam-Gyu Park for lending a Keithley 2400 to perform experiments.

**ToC FIGURE**

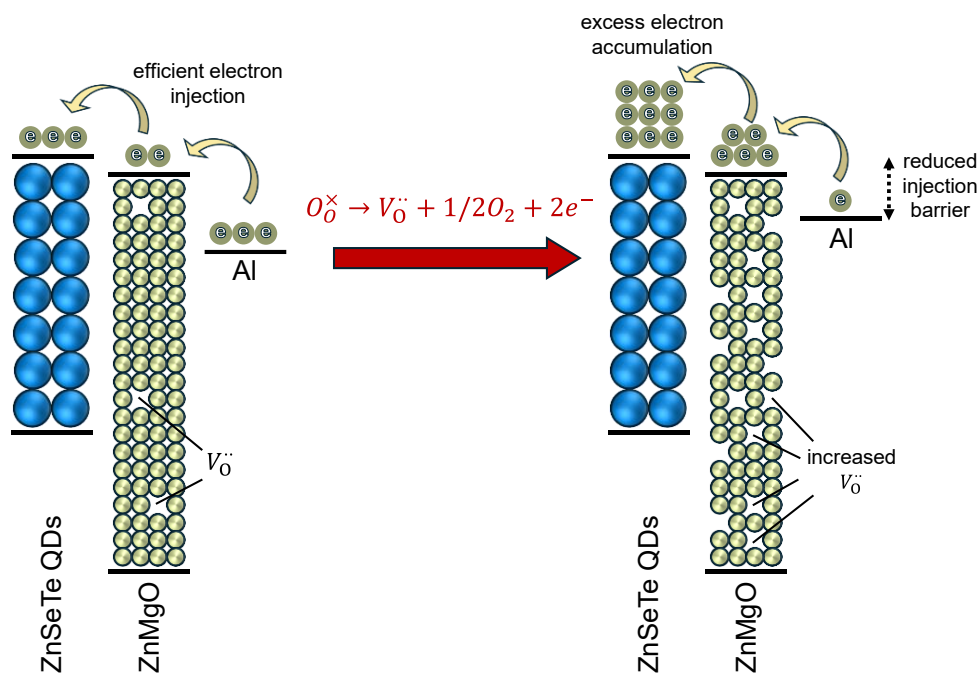

$O_O^\times \rightarrow V_O^{\cdot\cdot} + 1/2 O_2 + 2e^-$